# TopoTEM: A Python Package for Quantifying and Visualising Scanning Transmission Electron Microscopy Data of Polar Topologies

Eoghan N. O'Connell,[1][*][†] Kalani Moore,[1] Elora McFall,[1] Michael Hennessy,[1] Eoin Moynihan,[1] Ursel Bangert,[1] Michele Conroy[1,2][*]

1. Department of Physics, Bernal Institute, University of Limerick, Limerick, Ireland
2. Department of Materials, Royal School of Mines, Imperial College London, UK

*Authors for correspondence: eoghan.n.oconnell@gmail.com, m.conroy@imperial.ac.uk

**ABSTRACT:**

The exotic internal structure of polar topologies in multi-ferroic materials offers a rich landscape for materials science research. As the spatial scale of these entities are often sub-atomic in nature, aberration corrected transmission electron microscopy (TEM) is the ideal characterisation technique. Software to quantify and visualise the slight shifts in atomic placement within unit cells is of paramount importance due to the now routine acquisition of images at such resolution. In the previous ~decade since the commercialisation of aberration corrected TEM, many research groups have written their own code to visualise these polar entities. More recently, open access Python packages have been developed for the purpose of TEM atomic position quantification. Building on these packages, we introduce the TEMUL Toolkit: a Python package for analysis and visualisation of atomic resolution images. Here, we focus specifically on the TopoTEM module of the toolkit where we show an easy to follow, streamlined version of calculating the atomic displacements relative to the surrounding lattice and thus polarisation plotting. We hope this toolkit will benefit the rapidly expanding field of topology based nano-electronic and quantum materials research, and we invite the electron microscopy community to contribute to this open access project.

# INTRODUCTION:

The advent of commercialised aberration corrected TEMs and scanning electron probe TEMs (STEMs) directly corresponds with the initial rapid increase in interest for the new field of charged ferroelectric domain wall research (Borisevich et al., 2010; Jia et al., 2008; Seidel et al., 2009). Oxide based ferroelectric topologies were the ideal system to push the spatial resolution and showcase the use of aberration corrected STEM technology. Concurrently, the materials science research field could now experimentally prove previously predicted exotic topological states by theoretical calculations (Catalan et al., 2011; Jia, Urban, Alexe, Hesse, & Vrejoiu, 2011). Prior to the analysis of polar ferroelectric domain walls by aberration corrected STEM, it was accepted that unlike magnetic domain walls, ferroelectric walls were only one to two unit cells in width. However, this is in fact not the case in various ferroelectric material systems. There has been several unexpected polar and strain variations within and in the vicinity of domain walls quantified by STEM atomic displacement mapping.(Das et al., 2019; Hong et al., 2021) The advancements in these two research communities has always been symbiotic, with more recently new STEM techniques such as 4DSTEM being used to map out polarisation of complex higher order topologies such as skyrmions (Das et al., 2019) and merons(Shao et al., 2021). Additionally, ptychographic reconstructions of 4DSTEM data sets has allowed scientists to move beyond the spatial resolution possible by STEM annular dark and bright field detector imaging(Chen et al., 2021; Rothmann et al., 2020). In this paper we will focus on how to extract polarisation information from atomic resolution STEM data of polar ferroelectric and multiferroic topologies using post processing software.

Initially, the few groups that had access to corrected STEMs and collaborated with ferroelectric materials thin film growers developed code internally for the post-processing of the images. The cation displacement was quantified in the STEM images taken of domain wall topologies and thus polarisation could be assigned for each unit cell. However, there hasn't been a unified approach to this type of STEM data analysis. Our aim as a group of TEM and ferroelectric material scientists was to make a streamlined package that could easily be used across the various communities and more specifically by non-experts in code development. We hope that this open access software package will be developed by external users for different ferroelectric and multiferroic material systems of interest.

The TEMUL Toolkit (O'Connell, Hennessy, & Moynihan, 2021) is a Python image and data analysis library built on Atomap (Nord, Vullum, MacLaren, Tybell, & Holmestad, 2017), HyperSpy (de la Peña et al., 2021), Scikit-Image (Van der Walt et al., 2014) and other libraries. Our goal with the TopoTEM module of TEMUL Toolkit is to fill the niche of polarisation mapping and general electron microscopy analysis that may not fit into the aforementioned packages, allowing researchers to contribute code - from publications or otherwise - for others to use and expand on. We also aim to bridge the gap between packages for ease-of-use; for example, combining the atom finding capabilities of Atomap with the image simulation software of PyPrismatic (Ophus, 2017; Pryor, Ophus, & Miao, 2017) and abTEM (Madsen & Susi, 2021). Currently, the TEMUL Toolkit covers a broad spectrum of analysis from extensive polarisation visualisation to interactive image filtering, and basic line intensity profiling to element refinement (O'Connell et al., 2021). These functions and methods are described at www.temul-toolkit.readthedocs.io and in (O'Connell et al., 2021). A guiding principle of the package is dissemination through immediate reproduction of published data analysis for a scientific and general audience. This is accomplished by in-browser Jupyter Notebooks (Kluyver et al., 2016) run within a docker (Merkel, 2014) container provided by Binder (Jupyter et al., 2018) and requires no downloads. The majority of the TEMUL Toolkit package is built upon the data structures of HyperSpy and Atomap, specifically Atomap's sublattice class. It is therefore relatively easy for those familiar with these libraries to get started with the TEMUL Toolkit. Importing data from other languages, such as MATLAB, is also straightforward. The TEMUL Toolkit is open to any contribution, including those that lead to, or come from, published work.

The TopoTEM module of TEMUL Toolkit has already been used in several ferroelectric topology publications, including analysis of domain walls in PbTiO$_3$ (Moore, Conroy, O'Connell, et al., 2020) where the TopoTEM (formerly polarisation) and signal_plotting was used to quantify and visualise the changing polarisation. The curvature of SrRuO$_3$ interlayers of multi-stacked PbTiO$_3$ thin films was initially analysed by the combined use of Atomap and MATLAB code in the publication by (Hadjimichael et al., 2021). This was then fully adapted to Python within the TEMUL Toolkit and can now be accessed in the lattice_ structure_tools of the TopoTEM module. In both cases, the published data analysis can be verified and reproduced with a one-click Binder Jupyter Notebook running in the browser. In (Moore et al., 2021), where higher order vortex topologies were analysed, the TopoTEM module code was developed further to add in colour wheels and other tools for chiral transitions visualisation, as seen in the plot_polarisation_vectors function. Finally we have developed code to map out more complex unit cells distortions, and their resulting polarisations, such as in Boracite (Conroy et al., 2020). We hope this will give readers ideas to contribute code for unit cells of other polar ferroic materials systems.

In this paper we will focus on the polarisation calculation and visualisation tools available within the TopoTEM module of the TEMUL Toolkit. The TopoTEM polarisation module provides several functions for quick computing and visualisation of atomic column polarisation. The TEMUL Toolkit does not currently provide an exhaustive list of these computing methods, though contributions are welcome. First, we will describe some of the implemented methods of calculating the unit cell polarisation for various materials. We will describe the extensive *plot_polarisation_vectors* function and its various plotting parameters, as well as useful colour maps available within Python. We then discuss how to average the calculated polarisation over a number of squared unit cells to easily display the overall polarisation in an image. We also briefly describe the lattice structure tools available within TopoTEM. Finally, examples of the code applied to existing structures is presented.

**TopoTEM Module**

Here we describe several methods for calculating polarisation (atomic column shifts), all of which output the (x, y) coordinates and (**u**, **v**) vector components. These polarisation vectors can be visualised with the *plot_polarisation_vectors* function. Figure 1 shows an example polarisation mapping workflow.

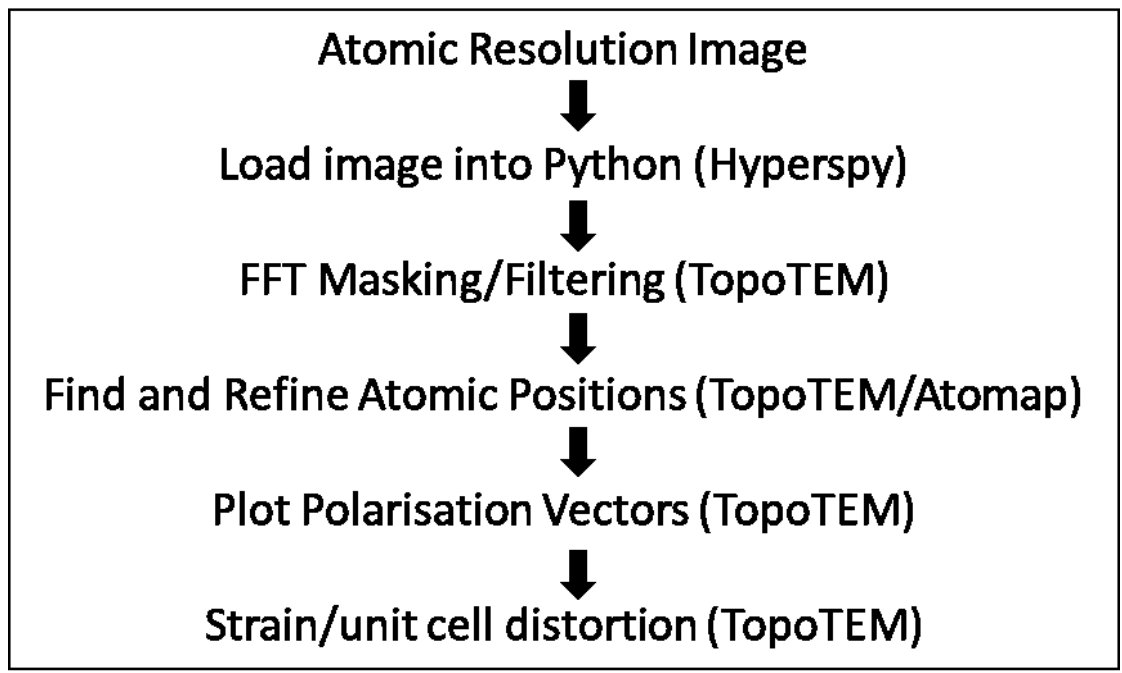

**Fig. 1.** Example analysis workflow for calculating and plotting polarisation vectors from atomic resolution images. The packages given in parentheses for each step are not exhaustive. Additionally, data output from other codes, e.g., MATLAB can be input to the final plotting step.

# **FFT masking and Filtering**

The temul.signal_processing module allows one to choose the masking coordinates with *choose_mask_coordinates* and returns the masked fast Fourier Transform (FFT) with get_masked_ifft. This can useful in various scenarios, from understanding the diffraction space spots and how they relate to the real space structure, to revealing domain walls and finding initial atom positions for complex unit cells or for materials where the first and second sublattice have similar z-contrast in the acquired STEM data.

**Fast Polarisation Mapping**

FFT masking has traditionally been used to quickly visualise the present domain patterns in TEM and STEM images of $ABO_3$ polar materials such as $PbTiO_3$, $BaTiO_3$ and $BiFeO_3$. Such $ABO_3$ perovskite ferroelectrics exhibit splitting of diffraction spots in different directions related to the different polar domains due to their strain and the non-centrosymmetric property (Moore, Conroy, & Bangert, 2020). We have integrated this into the work flow of the TopoTEM module, allowing users access to numerous data processing methods for calculating polarisation from STEM images in one package. Additionally we have integrated the function *get_masked_ifft*, which enables users to quickly monitor changes in polar topologies and associated domains for in-situ experiments and thus easily produce videos of such changes in STEM image series.

Another advantage of using this masking technique is that it can help find initial sublattice atom positions for complex unit cells or less than ideal real experimental TEM/STEM data. As shown in Figure 1, FFT masking can be used as an additional step to ease the next vital and computer memory heavy step of 'Find and Refine Atom Positions', and thus decrease the time spent finding correct atom positions. These steps are found in the 'Masked FFT and iFFT' of TEMUL Toolkit's TopoTEM module.

# Find and Refine Atomic Positions

## Plot and Visualisation of Subatomic Displacements

As summarised in Fig. *1* above, the atomic coordinates of all sublattices are located and refined during the initial atomic coordinate finding routine using Atomap (or another external program). This step is followed by calculation and plotting of polarisation vectors. This process has been applied to two simulation and one experimental structure in Figure 2. Example sublattices are shown in **Error! Reference source not found.**(a-c), with red and blue markers in Fig. *2***Error! Reference source not found.**(a-c) indicating the first (positive) and second (negative) ionic sublattice of their structures, respectively. Fig. *2*(d-f) displays the resulting calculated polarisation vectors of each structure in Fig. *2*(a-c), respectively. The methods for calculating these polarisation vectors are described in the following paragraph for the three example structures in Fig. *2*(a-c).

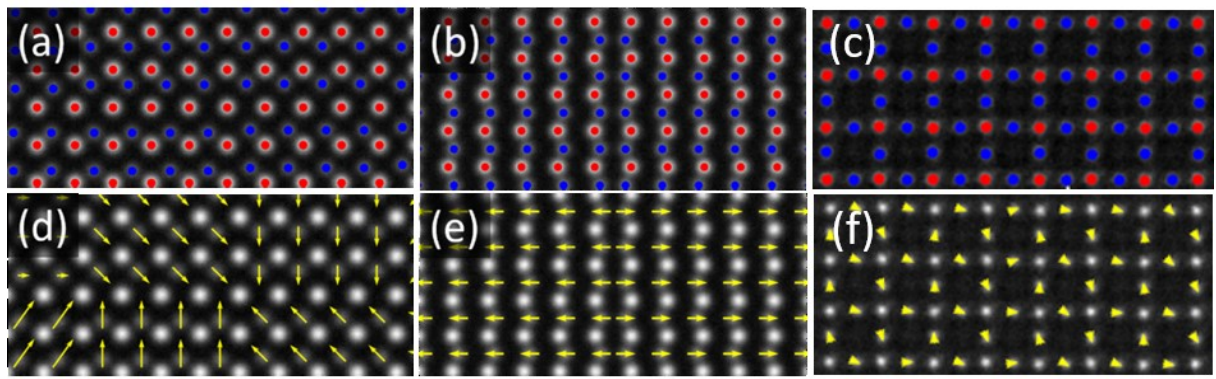

**Fig. 2.** Methods for calculating and displaying atomic displacements in atomic resolution STEM images. The red and blue markers in (a-c) indicate the position of the first (positive) and second (negative) ionic sublattice, respectively. (a) STEM simulation of perovskite $ABO_3$ polarised atomic structure ($PbTiO_3$). Atomap's *get_polarization_from_second_sublattice* function is useful and reliable for many such structures. For non-standard polarised structures as shown in (b) and (c), TopoTEM's *find_polarisation_vectors* function is used to calculate polarisation in such structures. (b) STEM simulation orthorhombic ferroelectric Cu, Cl boracite $Cu_3B_7O_{13}Cl$,(Conroy et al., 2020) (d-f) The resulting calculated direction and magnitude of the atomic displacement shift from the ideal structure of (a-c).

For classic perovskite $ABO_3$ polarisation, Atomap provides the *get_polarization_from_second_sublattice* function (described at www.atomap.org). Fig. *2*(a) shows this function in use for a standard atomic structure, such as $PbTiO_3$. The yellow vectors in Fig. *2*(d) are plotted using the *plot_polarisation_vectors* function. In this structure, they represent the reverse displacement of the second sublattice atomic columns (blue) relative to the centre of the surrounding four first sublattice atomic columns (red) (K. Moore, Conroy, et al. 2020).

For structures that vary from the $ABO_3$ [001] cubic projection the *plot_polarisation_vectors* function cannot be used to calculate polarisation. The example image in Figure 2(b) represents one such structure, wherein the polarisable atomic columns in the blue second sublattice are not located in the centre of the bright columns of red first sublattice, as in Figure 2(a). The initial and final coordinates of a sublattice of atomic columns can be used within TopoTEM's *find_polarisation_vectors* function, which calculates the u and v components of the vector. Therefore, the initial (ideal) coordinates and refined (actual) coordinates are input to the *find_polarisation_vectors* function to calculate the polarisation. The resulting polarisation vectors are displayed in Figure 2(e) using the *plot_polarisation_vectors* function.

Another example of a non-$ABO_3$ structure ferroelectric polar material is $Cu_3B_7O_{13}Cl$ (Cu, Cl based boracite) (McQuaid, Campbell, Whatmore, Kumar, & Gregg, 2017). This is an example of a polar material experimentally imaged by atomic resolution STEM for the first time, where initially we had to develop code to quantify and visualise the sub-atomic displacements within the TopoTEM module, as shown in (Conroy et al., 2020) and in Figure 2(c)&(f).

For each of the three methods from Figure 2(a-c), the resulting (x, y) coordinates and (u, v) components are output. The polarisation vectors can be input and plotted with the *plot_polarisation_vectors* function. Figure 2(d) shows the resulting polarisation vector due to the cation displacement of PbTiO$_3$ previously established in literature (Shirane, Axe, Harada, & Remeika, 1970; Yadav et al., 2016). Figure 2(e,f) plots the direction and magnitude of the atomic displacement shift from the ideal structure in Figure 2(b,c), respectively.

For some material systems the type of polarisation is uniaxial and thus in theory the polarisation can only be along one direction (e.g. no chirality). This is the case for structures such as uniaxial LiNbO$_3$ (Lee, Xu, Dierolf, Gopalan, & Phillpot, 2010). To calculate the polarisation in these materials, one uses the TopoTEM's *atom_deviation_from_straight_line_fit* function, adapted from Gonnissen et al. 2016. Figure 3(c) displays the atom plane directions via the *plot_planes* sublattice method. The kink in the atom planes present in the centre of Figure 3(c) indicates a domain wall topology. As described by Gonnissen *et al*, *n* atomic columns along the atom plane are fit to a straight line. This is repeated for the *n* atomic columns starting at the end of the atom plane (Gonnissen et al. 2016). The slope of the second fit is set equal to the slope of the first fit, while the y-intercept can vary. The line parallel and halfway between these two fittings is used as the line from which the **u** and **v** vector components are calculated from each atomic column. The direction of the plane of atoms, the order of fitting, as well as the number of atomic columns to use in the fitting of each plane of atoms, can be varied within the *atom_deviation_from_straight_line_fit* function.

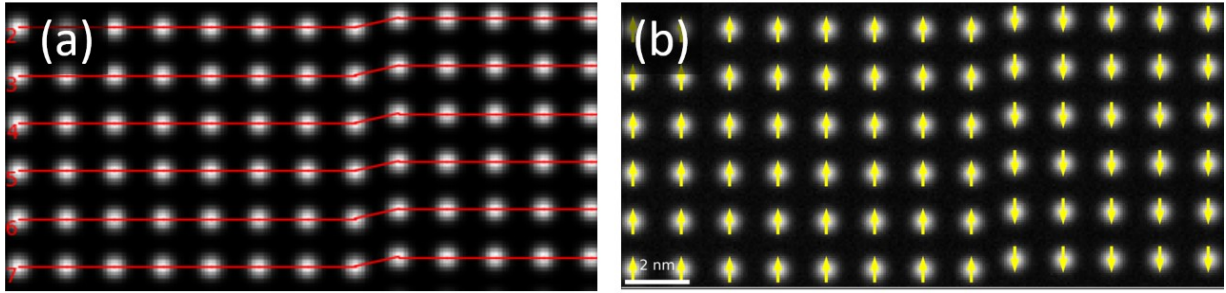

**Fig. 3.** Method for calculating polarisation in atomic resolution images of a uniaxial ferroelectric oxide. (a) STEM simulation of LiNbO$_3$ where the horizontal lines represent the atomic planes in a given direction, and (b) the resulting calculated polarisation vectors.

## **Plot Polarisation Vectors**

### **Visualising the Polarisation**

To display the calculated polarisation vectors, we use the flexible *plot_polarisation_vectors* function. This function is built on the popular Matplotlib Python library (Hunter 2007). It can be used to visualise calculated polarisation in several different styles by using the parameter *plot_style*. The *plot_style* parameter currently has five options to choose from: standard vector arrows, colour mapping, contour mapping, and two different colour wheel displays. Each of these is displayed in Figure 4. To display the polarisation vector arrows in a single colour, "vector" is input for the *plot_style* parameter as shown in Figure 4(a). Figure 4(b) displays the "colormap" option, and highlights the relative size of the vector magnitudes. Figure 4(c) shows the "contour" option for *plot_style*. Here, the contour colours represent the polarisation angle and are partly transparent to show the underlying atomic structure for reference. One may notice that the bottom half of 4(c) is a single colour, whereas it is represented by two colours in Figure 4(d-e). This is due to the categorical cyclic colour map used i.e., the green region only includes angles between 45 and 90 degrees. The "colorwheel" option presented in Figure 4(d) is useful for showing the polarisation vector angles at a glance. The "polar_colorwheel" option in Figure 4(e) communicates the relative vector magnitude as well as the vector angle.

Additionally, this allows users to use the same color scheme as the STEM differential phase contrast (DPC) color wheel plots used in commercial microscopy software such as Thermo Scientific Velox[TM] and JEOL 4DCanvas[TM].

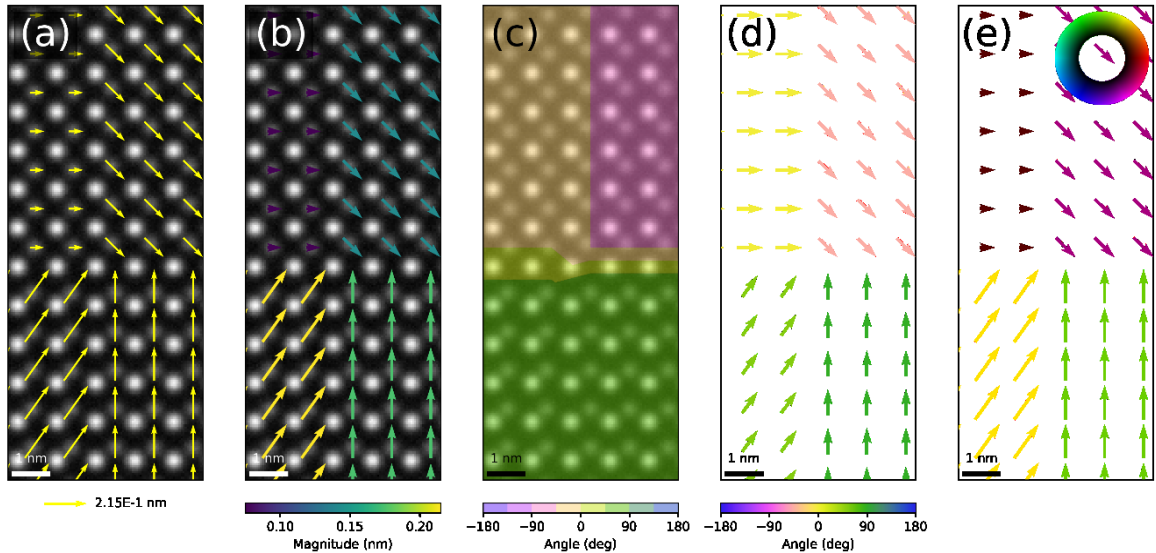

**Fig. 4.** Examples of plotting calculated polarisation with different plot_style inputs within the plot_polarisation_vectors function. (a) "vector", (b) "colormap", (c) "contour", (d) "colorwheel" and (e) "polar_colorwheel".

Apart from the *plot_style* parameter options described above, the *plot_polarisation_vectors* function has many parameters for modifying the plot. The polarisation data may be displayed with or without the background image, as in Figure 4(a) and 4(d), respectively. Figure 4(b) shows the "colormap" *plot_style* with a perpetually uniform colour map indicating the scaled magnitudes of the vectors. The angle of the polarisation vectors - which is often useful for visualising domains – is shown with the "colorwheel" *plot_style*, as displayed in Figure 4(c). An example of the "contour" *plot_style* overlaid on the image is shown in Figure 4(d), with the vector angle represented by the contour map. The *remove_vectors* parameter was used in this case to remove the vectors from the image. For the "colormap" and "contour" *plot_style* mapping, the *vector_rep* parameter is used to display either the magnitude or the angle of the polarisation vectors, as presented in Figure 4(b) and Figure 4(d), respectively. Setting all vectors to the same length is done by using the *unit_vector* parameter.

Utilising suitable colour maps allows for clear presentation of polarisation in an image. Colour maps are an essential part of the *plot_polarisation_vectors* function. For example in Figure 5 the colour map is used to clearly show the upward and downward polarisation as red and blue, respectively. This is much clearer than the yellow vector arrows as previously used in Figure 3(b) for the same material and domain wall topology. Using colour maps can remove distracting information in an image and allow the user or reader to focus on the polarisation trends. The colour maps used in each figure in this article are either taken from the Matplotlib package's perpetually uniform colour maps or the colour maps made available by the Colorcet package (which are also available through Matplotlib).

**Averaging Polarisation**

Atomic resolution polarisation of a structure can be averaged into regions defined by a unit area using the *get_average_polarisation_in_regions* function. The *get_divide_into* function is used to calculate a reasonable area for averaging, i.e., it will output the division needed for the desired unit cell averaging. The *get_average_polarisation_in_regions* function's (x, y, **u**, **v**) output can then be input to *plot_polarisation_vectors* as described previously. Example polarisation vectors averaged over 2x2 and 4x4 unit cells are shown in Figure 5(b) and Figure 5(c), respectively.

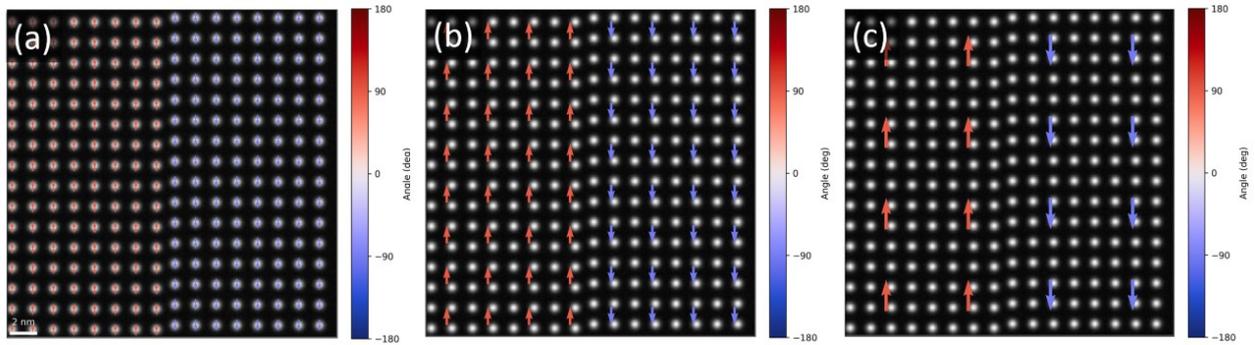

**Fig. 5.** Calculating polarisation in atomic resolution images of a uniaxial ferroelectric LiNbO$_3$ STEM HAADF simulation. (a) Polarisation colormap per unit cell at a domain wall topology, Polarisation averaged over (b) 2x2 and (c) 4x4 unit cells using the *get_divide_into* and *get_average_polarisation_in_regions* functions.

This function is especially useful for large images with many thousands or tens of thousands of unit cells. It allows for clear understanding of the overall polarisation in the image at a glance. In future, we hope that such an analysis routine could be used during acquisition (integrated into TEM vendor software) to quickly find interesting regions or domain walls in a specimen. An example of the visualisation and averaging of polarisation from an experimental STEM dataset to visualise a DW topology is shown in Figure 6.

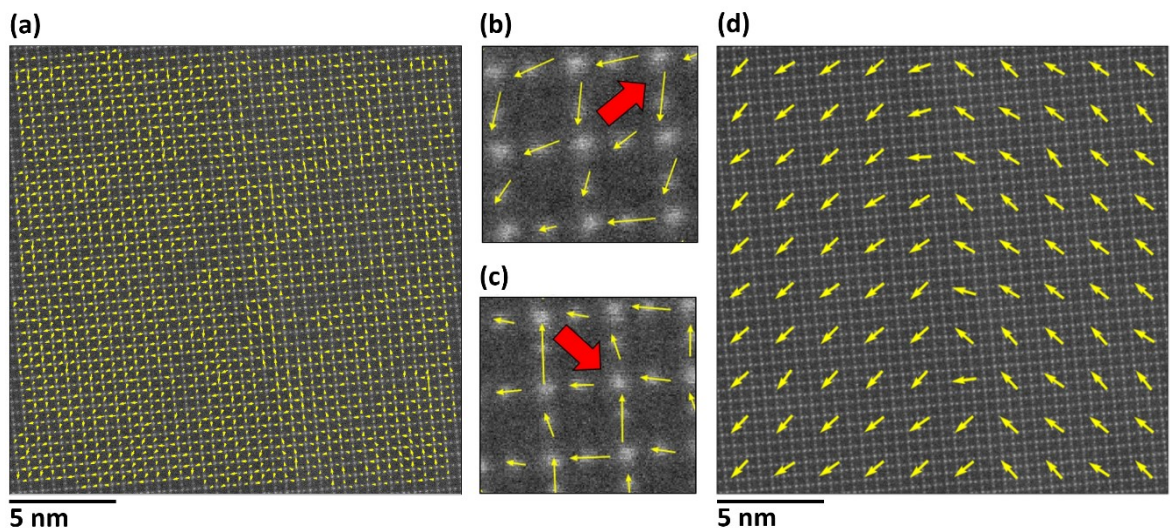

**Fig. 6.** Visualisation of polarisation in STEM image of trigonal ferroelectric Fe$_{2.4}$Mg$_{0.6}$B$_7$O$_{13}$Cl by plotting (a-c) individual cation polarisations and (d) averaged polarisation reveals the DW topology.

This also confirms that the *plot_polarisation_vectors function* and in general the TopoTEM data processing module is functionally robust to distinguish between the unit cell distortions of two different polar phases of the same material system (boracite in this case) from experimental STEM data. One can see clearly the difference between the polarisation plots of orthorhombic Cu-Cl boracite (Figure 2(f) and (Conroy et al., 2020)) and trigonal Fe, Mg – Cl boracite(Dowty & Clark, 1972) in Figure 6(b,c). Additionally, using the *get_average_polarisation_in_regions* function (Figure 6(d)) allows one to visualise the polarisation direction much more clearly compared to the non-averaged polarisation plotting in this large field of view and complicated unit cell material.

## **Strain/unit cell distortion**

### **Analysing Lattice Structure**

Strain engineering of ferroic materials can be used to form ferroelastic topologies and even induce ferroelectricity in nominally non-ferroelectric materials.(Choi et al., 2004; Haeni et al., 2004; Schlom et al., 2007) Several research groups have shown that with precise nanoscale control of the induced strain and electrostatic boundaries conditions, non-trivial polar topologies can be created (Hadjimichael et al., 2021; Tang et al., 2015; Wang et al., 2020; Yadav et al., 2016; Zubko et al., 2016). In additional to mapping out the cation and anion displacements within the unit cell, quantifying the unit cell distortion and the trends across topologies is vital to relate polarisation vector with strain.

The TopoTEM module includes tools for the analysis of lattice structure such as strain, curvature, atom plane angles, and c/a ratio of unit cells. In Moore et al. 2020 for STEM imaging of polar $PbTiO_3$ domain wall junctions using the *get_strain_map, rotation_of_atom_planes*, and *ratio_of_lattice_spacings* functions were used to quantify and visualise the strain, atom plane angles, and c/a ratios within the field of view. Example outputs from each of these functions are shown in Figure 7 and more information on the PbTiO$_3$ domain wall junctions can be found in (Moore et al. 2020).

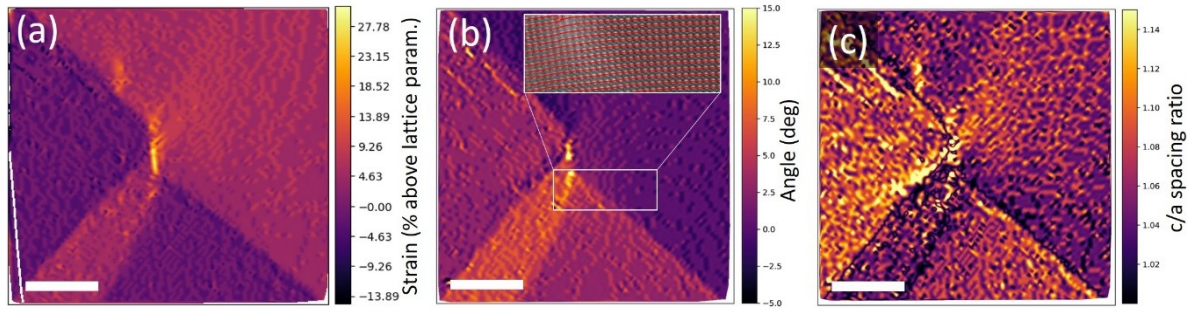

**Fig. 7.** Quantified (a) strain map, (b) atom plane angle map and (c) lattice spacing c/a ratio map of a PbTiO$_3$ domain wall junction described by (Moore et al. 2020). The horizontal atom plane structure is inset in (b) and shows the structure bending across the domain wall junction. Scale bars are 8 nm. Adapted from (O'Connell 2020).

The *ratio_of_lattice_spacings* function has also been used to calculate the shearing of unit cells in STEM imaging. For materials with such a small unit cell distortion due to structural shearing as for ferroelastic ferroelectric boracites (McQuaid et al., 2017; Zimmermann, Bollmann, & Schmid, 1970), strain mapping using geometric phase analysis algorithm (Hÿtch, Snoeck, & Kilaas, 1998; Strain++) from STEM data is not possible. Thus the *ratio_of_lattice_spacings* function provides a method to quickly check for strain coupled polar topologies present in materials, when diffraction techniques such as 4DSTEM are not possible or available. When viewing orthorhombic boracites in the [100] projection there are four in-plane and two out-of-plane domains that can be imaged. In Figure 8 the unit cell shearing and thus polarisation in-plane (xy) and out-of-plane (orthogonal to xy, along the beam axis) is measured across a domain wall. The atom to atom distance is aggregated (grouped) in 10 nm thick y axis slices and averaged to display the data in Figure 8(d). The X position (nm) on the x-axis can be directly compared with the x (nm) axis in Figure 8(b,c). From left to right, the grouped mean increases and decreases by the same amount for zone 2 and zone 3, respectively.

In the left, in-plane, domain there is a difference of ~15 pm between the zone 2 and 3 distortion. Moving toward the right, the grouped mean of zone 3 decreases by ~2 pm while the grouped mean of zone 2 increases by the same amount. This change in grouped mean indicates the location of the domain wall as ~15-30 nm along the x-axis. The in-plane/out-of-plane DW is inclined through the lamella, 45° to the viewing direction. So the transition from 15-30 nm reflects the gradual transition of the DW i.e., 0-15 nm (in-plane domain), 15-30 nm (mixed domain), and 30-80 nm (out-of-plane domain). The ~11 pm separation is therefore the unit cell distortion due to the ferroelastic primary order parameter in Cu-Cl boracite, because the domain is polarised out-of-plane.

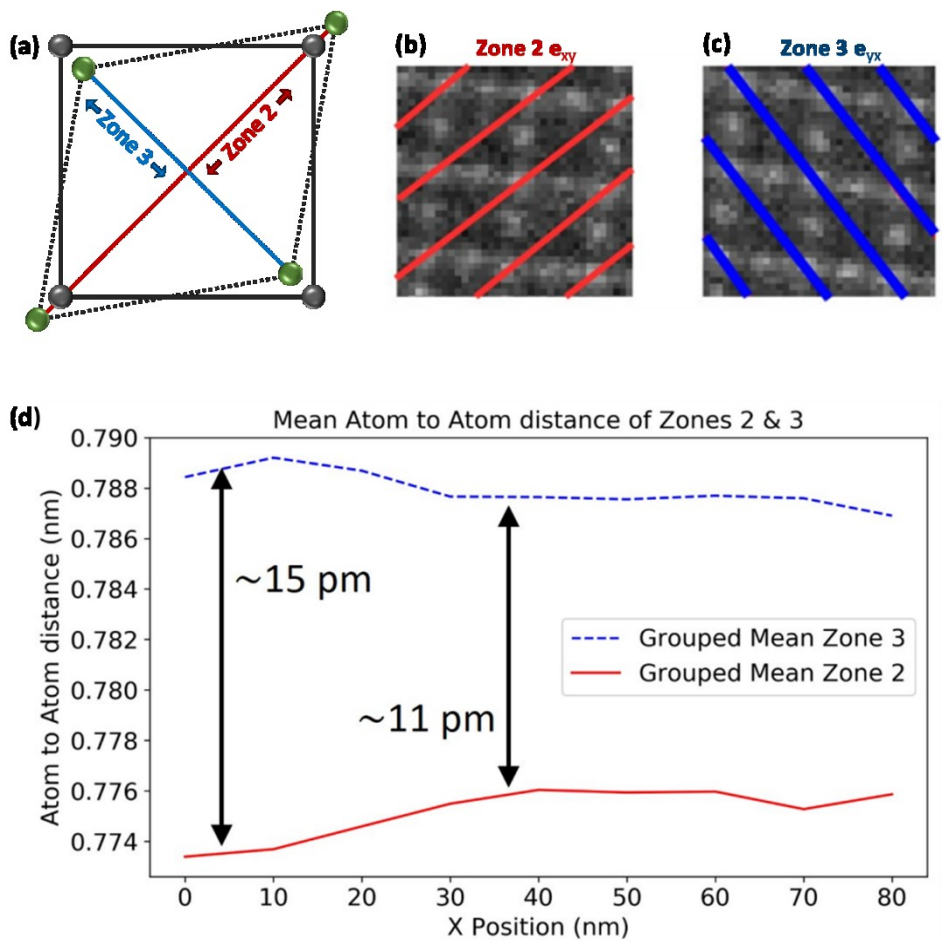

**Fig. 8.** (a) Schematic illustrating the shear distortion of orthorhombic boracites, (b) and (c) direction of the zone axes 2 and 3, from which the atom to atom distances are grouped and averaged, (d) plot across a DW of the average atom to atom distances in the zone 2 and zone 3 directions, and thus $e_{xy}$ and $e_{yx}$ shear strain plot.

Curvature strain engineering research of thin film ferroic oxides has produced several interesting polar topologies and phases results as shown recently by (Ji et al., 2019) and (Peng et al., 2020). A tool for calculating and visualising curvature in SrRuO$_3$ and PbTiO$_3$ thin films was adapted from MATLAB code used in (Hadjimichael et al., 2021) and is now accessible in the *calculate_atom_plane_curvature* function. An example use of this code on the data from Hadjimichael et al., is displayed in Figure 9.

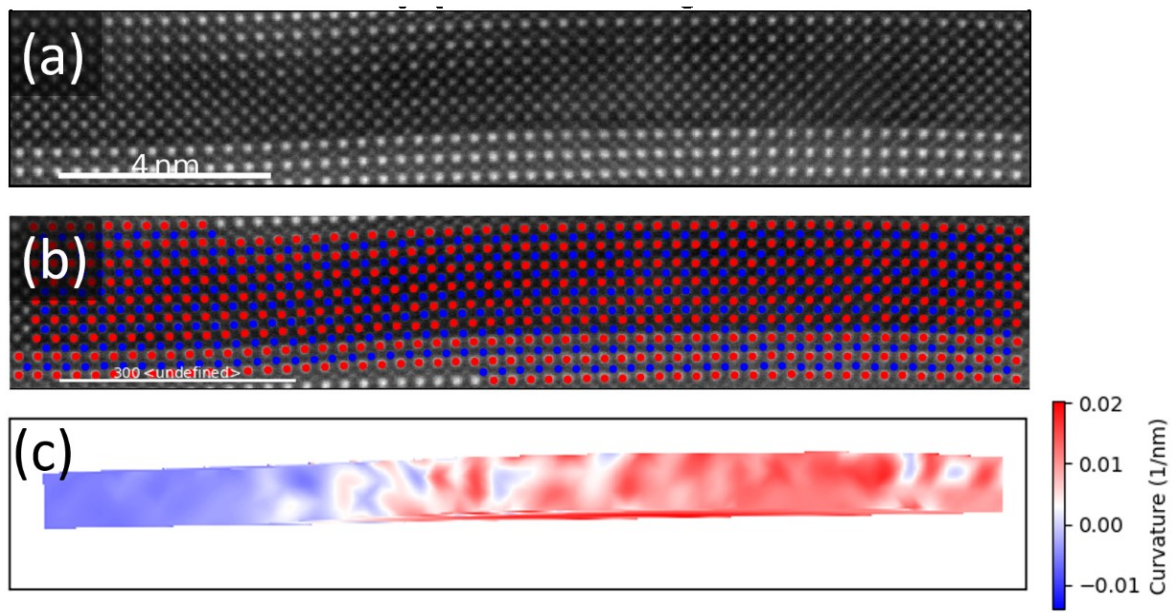

**Fig. 9. (a)** STEM ADF image of PbTiO$_3$–SrRuO$_3$- PbTiO$_3$ interfaces **(b)** atomic position overlaid on image **(a)**, and **(c)** quantified curvature map of the SrRuO$_3$ layer.

# **Conclusion**

The TopoTEM software module allows users to easily compute and visualise polarisation in atomic resolution STEM images all in one package. By building on the HyperSpy and Atomap packages, among others, we have created a streamlined workflow to detect polar topologies in traditional ferroelectric materials such as $PbTiO_3$, $LiNbO_3$, and materials with more complex unit cells and atomic displacements such as boracites and aurivillius phases. The *get_average_polarisation_in_regions* extends this plotting by providing an easy method for computing the average polarisation in large areas, and therefore the user can quickly identify polar regions of interest and polarisation pattern trends within scans. TopoTEM also includes several lattice structure tools allowing one to directly correlate polarisation and strain per unit cell. All of the above functionality has been applied to ideal STEM simulations and more complex structures of experimentally collected STEM data. We will continue to develop this open access package adding polarisation plot tools for different material systems and crystallographic projections. We believe that our software package will find widespread application in both the polar materials community and the atomic resolution STEM community at large.

**Acknowledgements**

E.O.C was supported by the Irish Research Council under the Postgraduate Government of Ireland grant GOIPG/2015/2410. M.C acknowledges funding from Science Foundation Ireland Industry Fellowship (18/IF/6282) and Royal Society Tata University Research Fellowship (URF\R1\201318). M.H. was supported by VolkswagenStiftung (grant no. A123203). E.M. was supported by the Irish Research Council Enterprise Scheme (EPSPG/2017/311), in conjunction with the Ernst Ruska-Centre, Forschungszentrum Jülich. E.O.C., K.M., U.B., and M.C. acknowledge funding from the US-Ireland R&D Partnership Programme (grant no. USI 120) and Science Foundation Ireland (16/US/3344).

**Present Addresses**

† Max Planck Institute for the Science of Light & Max-Planck-Zentrum für Physik und Medizin, Erlangen, Germany